# Electrically controlled propulsion of skyrmions in chiral nematic

Oleg D. Lavrentovich

Advanced Materials and Liquid Crystal Institute, Department of Physics, Kent State University, Kent, OH 44242 USA

* Correspondence: olavrent@kent.edu

**Design of soft matter capable of controllable microscale dynamics is a frontier of modern science. Jiahao Chen et al. demonstrate that an electric field can drive particle-like solitons-skyrmions in a chiral nematic along a preprogrammed trajectory with a variable speed within a two-dimensional plane. The effect is rooted in flexoelectric polarization of a deformed director field.**

Dynamics of small particles in fluids has fascinated scientists for centuries, starting with the discoveries of self-propelled bacteria by Leeuwenhoek and chaotic motion of pollen grains in water by Brown. Controlling such dynamics is challenging because at the microscale, viscous forces overcome inertia, thus slowing down any motion, while thermal fluctuations disrupt organized flows. To overcome these difficulties, scientists develop approaches to activate motion by an external electric field, and to design the carrier medium and microparticles. Usually, the microscale dynamics is explored for an isotropic environment, such as water, a natural habitat of microorganisms. Technological advances, however, do not need to be limited by isotropic fluids. Recent years have seen a growth of interest in anisotropic fluids, such as a nematic liquid crystal, as a means to control microscale dynamics.

In a nematic, rod-like molecules align parallel to each other, along a so-called director, $\hat{\mathbf{n}}$. The orientational order brings about features of importance for microscale dynamics. First, it imparts a sense of direction as motion along $\hat{\mathbf{n}}$ faces a weaker viscous drag than motion perpendicular to $\hat{\mathbf{n}}$ [1]. Second, director reorientations trigger hydrodynamic flows. Third, spatial variations of $\hat{\mathbf{n}}$ produce topological defects that often behave as material particles and can be moved by external fields [2]. However, most nematics are paraelectric and do not distinguish between the heads and tails of the molecules, even when they are different. Thus $\hat{\mathbf{n}}$ and $-\hat{\mathbf{n}}$ describe the same state, $\hat{\mathbf{n}} \equiv -\hat{\mathbf{n}}$. Furthermore, nematics permit only point- and line-like topological defects, while the most interesting defects are three-dimensional particle-like solitons, representing director distortions confined within a finite volume $V$.

Imagine such a soliton of size $D$. A decrease in size $D \to D/\mu$, by a factor $\mu > 1$, increases the elastic energy density $(\nabla \hat{\mathbf{n}})^2 \propto 1/D^2$ by a factor $\mu^2$, while the volume $V$ decreases by a factor $1/\mu^3$. Therefore, a shrinking soliton lowers the total elastic energy, $\int (\nabla \hat{\mathbf{n}})^2 dV \propto 1/\mu$, and must disappear [2].

Particle-like solitons can be stabilized if a nematic is doped with chiral molecules, which force the director to twist in space, usually with a microscale pitch. Regions $\hat{\mathbf{n}}$ and $-\hat{\mathbf{n}}$ of a chiral nematic are separated by half of the pitch and cannot be moved closer to each other without a dramatic increase of the elastic energy. A similar stabilization, called Dzyaloshinskii-Moriya mechanism, is known in ferromagnets [3]. Chiral nematics support a rich plethora of particle-like solitons, including the so-called skyrmions, as reviewed recently by Wu and Smalyukh [4].

Writing in *Newton*, Bingxiang Li, Rui Zhang, Juan J. de Pablo, and Yangqing Lu and colleague from Nanjing University of Posts and Telecommunications, Nanjing University, The Hong Kong University of Science and Technology, University of Chicago and Argonne National Laboratory, demonstrate an exciting electric field control of dynamic particle-like skyrmions in chiral nematics [5]. The liquid crystal is confined between two glass plates with transparent electrodes. Surface interactions set the director parallel to the plates, say, along the $x$-axis, $\hat{\mathbf{n}} = (1,0,0)$. An alternative-current (AC) electric field realigns $\hat{\mathbf{n}}$ parallel to the $z$-axis normal to the plates. The realignment is caused by the positive dielectric anisotropy $\Delta \varepsilon > 0$ as the permittivity along $\hat{\mathbf{n}}$ is higher than the permittivity along perpendicular directions. The dielectric realigning torque is proportional to the square of field. Sometimes the two equivalent states, $\hat{\mathbf{n}}$ and $-\hat{\mathbf{n}}$, find themselves separated by a $\pi$ twist of $\hat{\mathbf{n}}$. A skyrmion is comprised of one of these states in the center and another at the periphery, connected by a twist that propagates radially in the plane of the cell. Its shrinking is impossible as explained above. The researchers noticed that when a direct current (DC) field supplements the AC one, then the skyrmion can be steered along any direction, by changing the amplitude and polarity of the DC field.

The observation poses an intriguing question of how the nearly axially-symmetric skyrmion acquires the sense of direction. The dependency on the field polarity is also puzzling since the dielectric torque is not sensitive to the field polarity, as established in numerous electro-optical studies that culminated in the industry of liquid crystal displays [6]. The researchers propose that the sense of direction is enabled by the so-called flexoelectric effect predicted by Robert B. Meyer [7]. Meyer considered nematic molecules with dipoles and a conical shape. In a uniform nematic, the probability of cones to point to the left and to the right is the same, thus there is no net polarization, Fig.1a. However, if the director experiences a splay, then the parity is broken, Fig.1b, producing a macroscopic electric polarization $\mathbf{P}_{fl}$, Fig.1b. Mathematically [7],

$$\mathbf{P}_{fl} = e_1 \hat{\mathbf{n}} \operatorname{div} \hat{\mathbf{n}} - e_3 \hat{\mathbf{n}} \times \operatorname{curl} \hat{\mathbf{n}}, \tag{1}$$

i.e., $\mathbf{P}_{fl}$ relates to the splay vector $\mathbf{s} = \hat{\mathbf{n}} \operatorname{div} \hat{\mathbf{n}}$ and the bend vector $\mathbf{b} = \hat{\mathbf{n}} \times \operatorname{curl} \hat{\mathbf{n}}$; $e_1$ and $e_3$ are flexoelectric coefficients. These two vectors are polar, $\mathbf{s} \neq -\mathbf{s}$ and $\mathbf{b} \neq -\mathbf{b}$, but still preserve the underlying $\hat{\mathbf{n}} \equiv -\hat{\mathbf{n}}$ symmetry. More symmetric quadrupolar molecules, without permanent electric dipoles, also produce flexopolarization, with $e_1 = e_3$ in Eq.1. [8].

Qualitatively, the directional skyrmion propulsion is rooted in the coupling of $\mathbf{P}_{fl}$ to the electric field, which produces a torque linear in the field, $\mathbf{\Gamma}_{fl} = \mathbf{P}_{fl} \times \mathbf{E}$. The AC component causes oscillation of $\mathbf{P}_{fl}$ and $\hat{\mathbf{n}}$. In its turn, the oscillation creates hydrodynamic flows that guide the locomotion along a time averaged $\mathbf{P}_{fl}$. The DC component affects symmetry of director deformations and thus the amplitude and direction of $\mathbf{P}_{fl}$, acting as a steering wheel.

Flexoelectricity-enabled propulsion of skyrmions involves the AC and DC fields, bulk elasticity and surface anchoring, anisotropic viscosity, etc., and thus requires a sophisticated numerical analysis [5]. However, the underlying principle can be illustrated by a simpler example of dissipative solitons called directrons [9], Fig.1c-f. The directrons form in a uniform nematic, $\hat{\mathbf{n}}_0 = (0,1,0)$ (Fig.1c), as localized perturbations $\hat{\mathbf{n}}(\mathbf{r})$ with a flexoelectric polarization $\mathbf{P}_{fl}$, excited by an AC field; Fig.1d shows a splay directron and Fig.1e shows a bend directron. The AC electric field causes up and down oscillations of $\mathbf{P}_{fl}$ (Fig.1f). These oscillations create hydrodynamic flows with direction that depends on the direction of $\mathbf{P}_{fl}$. For example, the splay directron propels along $\hat{\mathbf{n}}_0$, Fig.1d, while the bend one propels perpendicularly to $\hat{\mathbf{n}}_0$ (Fig.1e,f). The same principle, applied to a more complex geometry of skyrmions, is operational in Ref.[5].

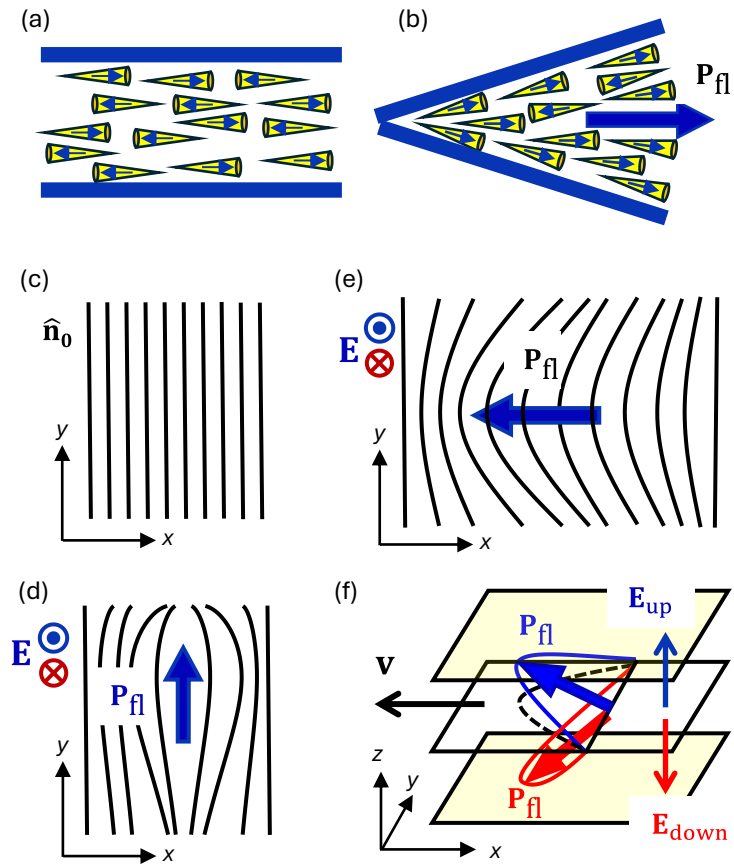

Figure 1: Flexoelectricity-enabled locomotion of directrons. a) A uniform paraelectric nematic comprised of cone-like molecules shows no electric polarization in the ground state. b) A macroscopic flexoelectric polarization $\mathbf{P}_{fl}$ emerges in a nematic with a splay deformation. c) A nematic is uniformly aligned when there is no external electric field. d), e). An AC electric field enhances a fluctuative splay, $\mathbf{s} = \hat{\mathbf{n}}\,\mathrm{div}\hat{\mathbf{n}}$, and bend, $\mathbf{b} = \hat{\mathbf{n}} \times \mathrm{curl}\hat{\mathbf{n}}$, respectively, thus creating dissipative solitons-directrons. f) The AC field causes oscillations or $\mathbf{P}_{fl}$ and propagation of a bend directron with a velocity $\mathbf{v}$. Panels c)-f) are redrawn with modifications from Ref.[2].

The discovered omnidirectional dynamics of particle-like skyrmions in chiral nematics adds a new dimension to the electrically controlled microscale hydrodynamics of soft matter. It might lead to applications such as targeted delivery of optical information and micro-cargo as the deformed director regions attract colloidal inclusions [10]. Similar effects could be observed in recently discovered flexoelectric nematics, in which the polarization is present even in ground state and couples linearly to an electric field. Collective dynamics of skyrmions would be another interesting avenue to explore, as these would interact in a nematic not only hydrodynamically but also through elastic and electrostatic forces.

**Acknowledgements**

The research of the author is supported by NSF grant DMR-2341830 and ARO MURI grant 84914-SM-MUR.

**Declaration of Interest**
The author declares no competing interests.